# CsPbBr(Cl,I)$_3$ quantum dots in fluorophosphate glasses

Kolobkova E.V.[1,2*], Kuznetsova M.S.[3], Nikonorov N.V.[1]

[1]Research Center for Optical Materials Science, ITMO University, Birzhevaya line 4, St. Petersburg, 199034, Russia

[2]St. Petersburg State Institute of Technology (Technical University), Moskovsky pr. 26, St. Petersburg, 190013, Russia

[3]Spin Optics Laboratory, Saint Petersburg State University, 198504 St. Petersburg, Russia

*kolobok106@rambler.ru

*Abstract*

For the first time the quantum dots CsPbX$_3$ (X=Cl, Br, I) in the fluorophosphate glasses were prepared. The samples were precipitated by two methods: (i) through self-crystallization during cooling of the glass melt and (ii) heat treatment of the glass. Controlled formation of CsPbX$_3$ quantum dots was realized by adjustment of cooling rate and heat-treatment conditions. The X-ray diffraction data was confirmed CsPbCl$_3$(Br$_3$, I$_3$) quantum dots formation. It was shown that, CsPbX$_3$ (X=Cl, I) quantum dots are formed in a cubic modification, while CsPbBr$_3$ in orthorhombic one. .The photoluminescence of quantum dots have high intensity with quantum yield 45-50 % and narrow band emission. The combined introduction of two anions (Cl/Br and Br/I) led to the simultaneous formation of two types of quantum dots, and indicates the difficulty of the anion exchange

## 1. Introduction

Optical materials based on the cesium lead halide perovskite nanocrystals are perspective ones because of their unique optical [1–3], optoelectronic [4], and photovoltaic [5] properties. Therefore a formation of CsPbX$_3$(X=Cl, Br,I) perovskite quantum dots (QDs) and nanocrystals (NCs) has become a top challenge in optical materials science.Application prospects of CsPbX$_3$ QDs have been demonstrated on the lasers [6], polarizers [7], light-emitting diodes [8], solar cells[9], and photodetectors[10].

Controlled changing of the QDs sizes or their compositions using anion and/or cation exchange [2] can modify the spectral properties of lead halide perovskite NCs. Ion exchangeis more frequently adopted for colloidal prepared nanocrystals, primarily because of the ease of ion exchange for lead halide perovskites, making the quantum confinement effect seemingly redundant for colloidal perovskite NCs. A comparison of conventional cadmium sulfoselenide QDs, which require considerable size tuning in order to cover the full visible spectrum range, with colloidal $CsPbX_3$ NCs demonstrates some advantages of the latter. The advantages of colloidal NCs perovskites in comparison with conventional cadmium sulfoselenide QDs include the tolerance of their luminescent parameters to surface defects and significantly easier fabrication methods [11].

In recent years, a large number of studies were devoted to the synthesis and properties of colloidal perovskite quantum dots (PQDs) [1–14]. However, the main disadvantage of these PQDs is low resistance to the environment and high temperatures. To stabilize CsPbX3, various types of protective coatings were created. [15-18]. Incorporation of perovskite NCs into polymers mesoporous materials (such as silica, $Al_2O_3$, $CaF_2$ and $TiO_2$) have been shown effective to improve the water resistance and stability. However, these coating are not sufficient to protect perovskiteNCs from high temperature. Therefore, new approaches toward more stable $CsPbX_3$ perovskite NCs are still highly required. One of such approaches may be the formation of nanocrystals and quantum dots in a glassy optical matrix by diffusion-controlled phase decomposition.

Nowadays a number of works on the synthesis of perovskites in glasses have been published [19-26]. These works were devoted to precipitation of the QDs by conventional way based on the crystallization of the initial glass during long exposure above the glass transition temperature. Quantum dots and NCs have been synthesized in borosilicate [19-21, 25. 26], phosphate with silicate additives [22], and borogermanate [23] glasses. It has also been found that PQDs into the glasses are not agglomerated, and maintain the chemical and optical stability. Lead-halide perovskite NCs in glasses can become attractive luminescent materials for many applications due to their strong light absorption and high quantum yield (QY), as colloidal nanocrystals. Lead-halide perovskite NCs luminescence is characterized by narrow monochromatic bands, which due to a change in the anionic components in the chlorine-bromine - iodine system covers the entire visible range due to sequential shift. This opportunity arises from change in band gap energies of $CsPbX_3$NCs depending on the X [cubic form of $CsPbCl_3$: $E_g$=2.82 eV (440 nm) and orthorhombic $CsPbBr_3$: $E_g$=2.36 eV (525 nm) as well as form of $CsPbI_3$: $E_g$=1.72 eV (720 nm)].

In the previously studied glasses, the shift of the luminescence band was carried out mainly, as in the case of colloidal perovskites, due to ion substitution, rather than by successive changes in the size of the NCs. In [23-26] the synthesis of solid solutions of perovskites was considered when replacing bromine with chlorine and iodine. It was concluded that in glasses, as in colloidal solutions, the anion substitution reaction was successful. The original glasses were made by a facile melting-quenching technique of CsPbX$_3$ (X = Br, I) QDs. The glasses demonstrated a tunable photoluminescence (480 –698 nm) controlled by the molar ratios of halide precursors. It was found that anion exchange reactions did not transform the crystal phase of the NCs during glass heat treatment and the patterns collected on the exchanged of CsPbX$_3$ (X = Br, I) QDs were in good agreement with those recorded on directly synthesized colloidal CsPbI$_3$ and CsPbBr$_3$ NCs.

The choice of fluorophosphates (FP) glassy matrix for formation of CsPbX$_3$ (X = Cl, Br, I) QDs in our study was at the first time due to the possibility of introducing high concentration of halides. Our previous studies have shown the versatility of the fluorophosphates matrix used in photonics. An important point was also the use of the previously developed synthesis technique [27], which allows keeping in the glass controlled active components content, in particular bromine, chlorine and iodine [29]. The using of fluorophosphate glass had made a possibility to synthesize lead sulfoselenide QDs [27], cadmium sulfoselenide QDs [28], halide copper nanocrystals [29] and achieve unique optical characteristics. Also the possibility of Ag$^+$/ Na$^+$ ion exchange and obtaining an optical waveguide with the subsequent formation of silver nanoparticles in the bulk and on the glass surface was shown [30, 31]. Thus, the high chemical resistance of the glasses under study was demonstrated not only to environment, but also to molten salts at temperatures above 300 ° C.

In present study, CsPbX$_3$ QDs with tunable visible emission were precipitated in fluorophosphate glasses through conventional melt-quenching and heat-treatment method. Absorption and photoluminescence (PL) spectra, X-ray diffraction(XRD) patterns all showed that CsPbX$_3$ QDs have been formed in the glass.

## 2       Experimental procedures

### 2.1    Glass ceramic preparation

Sample of the fluorophosphates (FP) glass with composition 40P$_2$O$_5$–35BaO-5NaF (NaCl)-10AlF$_3$-5Ga$_2$O$_3$–2Cs$_2$O–2PbF$_2$-1BaBr(I)$_2$, (mol. %) were elaborated using melt-quench technique.

The glass synthesis was performed in closed glassy carbon crucible at temperature, T=1000 °C. About 50 g of the batch was melted in a crucible for 20 min. Than the glass melt was cast on a glassy carbon plate and pressed to form a plate with thickness ~2 mm.  QDs were precipitated by the two different methods, including glass crystallization via  glass heat-treatment (1) and the glass self-crystallization during melt-quenching and further growth during heat treatment (2). The first method is a conventional technique for glass ceramic preparation. On the first stage, we obtain the initial glass, which has no color and accordingly does not contain QDs. QDs are obtained during secondary heat treatment at temperatures above the glass transition temperature. From the second method, QDs were obtained during the cooling of the melt by controlling of the residual concentration of bromine, chlorine or iodineions and the cooling rate of the glass melt. Unfortunately, the first method does not allow obtaining the content of quantum dots sufficient for analysis based on the X-ray method. Therefore, in the further study we will use the samples obtained by the second method.A much more pronounced diffraction signal was found in the XDR pattern of the sample prepared from the second method, ascribing to the higher halides content.

**2.2    Glass characterization**

Differentials canning calorimeter STA 449F1 Jupiter Nietzsche was used for measurement of the glass transition temperature (Tg). Transition temperature of the glass matrix was $T_g$=400°C. Nabertherm muffle furnace with program control was used for the thermal treatment of the samples at air atmosphere. Samples were placed into the furnace heated up to temperature required and, after an exposure, extracted from the hot furnace into air. Specimens were treated in the temperature region 400- 450 ºC within 20-40 minutes. The precipitation of QDs from glass was identified by X-ray diffraction. Cu-K irradiation ($\lambda$= 1.5406 A) with ascanning rate of 2°/min was used for the measurement with aresolution of 0.02° using a Rigaku X-ray diffractometer

The absorption spectra of FP glass samples under study were recorded in the 200- 800 nm spectral region using Lambda 650 Perkin Elmer spectrophotometer. The registration of luminescence properties was carried outby MPF-44A (Perkin-Elmer) spectrofluorimeter and Absolute PLQuantum Yield Measurement System (Hamamatsu).

**3    Experimental results and discussion**

**3.1    XRD study**

The X-ray diffraction patterns of the glasses prepared by second method were shown in Figure 1-3. Several weak diffraction peaks were observed, indicating the formation of crystalline phases in the glasses.

An analysis of the X-ray diffraction spectrum of glasses containing nanocrystals showed that when only the chloride component was introduced, NaCl was precipitated as a crystalline phase due to the presence of sodium ions in the initial glass. It was found that small additions of bromide ($BaBr_2$) should be added to the glass to obtain $CsPbCl_3$. Despite the presence of bromine in the composition, the diffraction peaks of nanocrystals in the glasses under study completely correspond to CsPbCl$_3$ (Fig. 1). It can be assumed that heterogeneous crystallization occurs, and CsPbBr$_3$ nanocrystals are formed as a nucleus with subsequent growth of CsPbCl$_3$ on its surface. Another option is a formation of mixed perovskites with a small admixture of bromine, which does not change the position of diffraction peaks. X-ray diffraction patterns of glasses doped with CsPbCl$_3$ nanocrystals are shown in Figure 1. Figure 1 shows absence of the double peaks at double angle of diffraction 2θ~ 30 °, as a key feature of the orthorhombic phase compared to the cubic phase. Thus, we can conclude that the cubic phase is formed. The estimation of the CsPbCl$_3$ NCs size using the Debay-Sherer equation gives a size of 7.0 nm (insert Fig. 1), which is significantly larger than the exciton Bohr radius of CsPbCl$_3$ a$_B$=4.0 nm. Earlier, when considering copper halide nanocrystals in FP glass, we have found that nanocrystals were normalized in a high-temperature hexagonal modification, which were preserved after the glass cooling to room temperature [28]. We have associated this effect with the influence of the glass matrix. The crystal size of 9.0 nm was revealed to be critical, below which the CuBr nanocrystals possess cubic structure, and above which the structure of the nanocrystals becomes hexagonal [29]. A preliminary explanation of the effects revealed is proposed, assuming the principal role of internal pressure in drops of a copper halide liquid in a glass matrix. The resulting picture can be extended to perovskite NCs: large NCs were normalized in high-temperature phase.

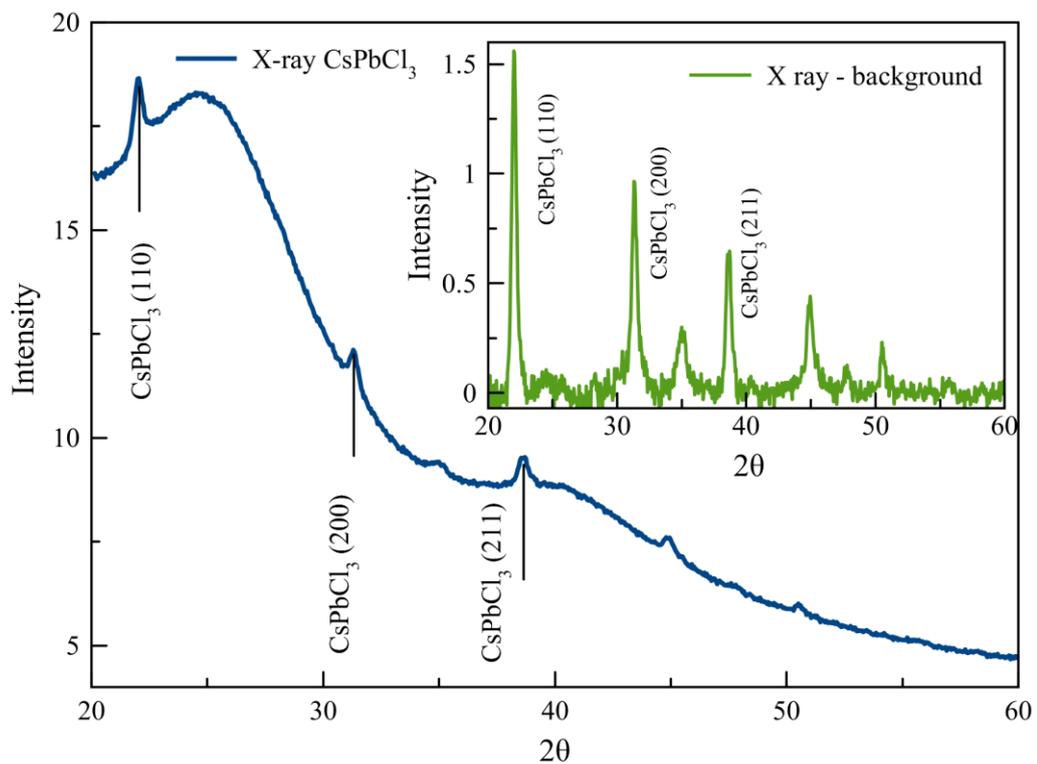

Figure1    The X-ray diffraction patterns of the glasses doped with CsPbCl$_3$(JCPDS No.73-0692); Inset shows X-ray diffraction patterns after subtraction the background.

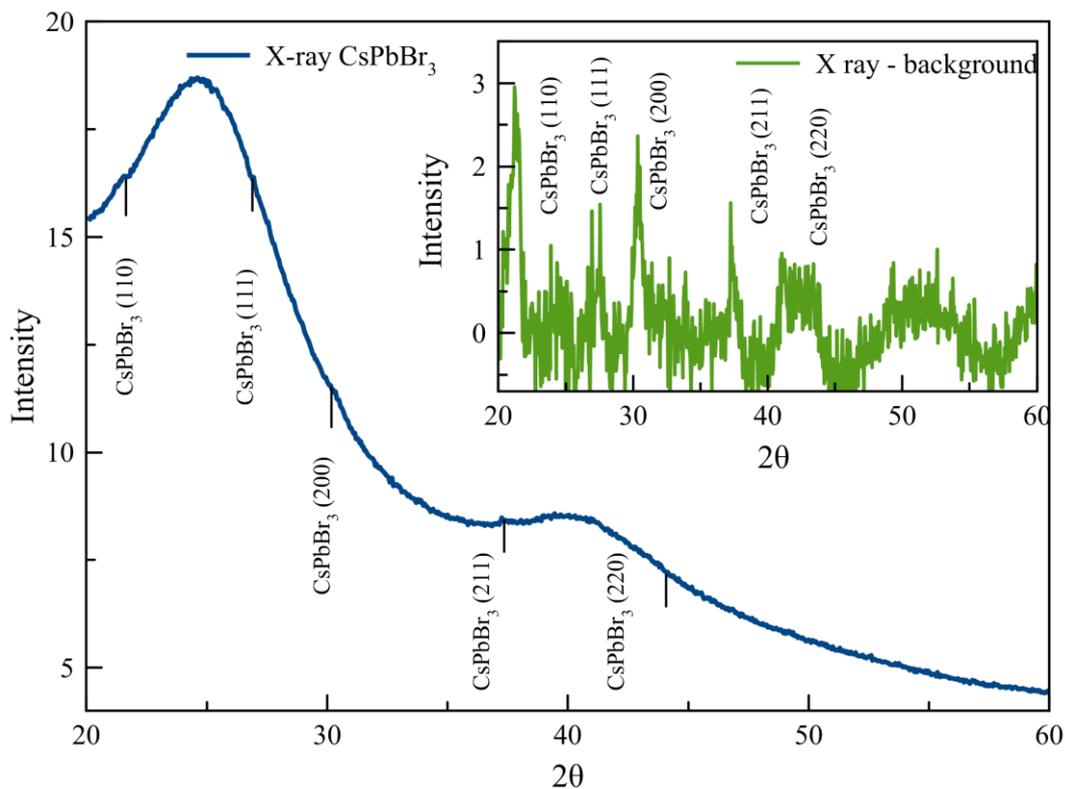

Figure 2    X-ray diffraction patterns of the glasses doped with CsPbBr$_3$ (JCPDS No. 75-0412); Inset shows X-ray diffraction patterns after subtraction the background.

Figure 2 shows of the X-ray diffraction patterns of the glasses doped with CsPbBr$_3$ NCs. Initial research suggested the structure of CsPbBr$_3$ nanocrystals (NCs) as cubic, but further studies revealed that both bulk CsPbBr$_3$ [35] and CsPbBr$_3$ NCs [36] are actually orthorhombic. We note that, for small CsPbBr$_3$ NCs with broadened XRD peaks, it is difficult to distinguish between cubic and orthorhombic phases both exhibiting similar Bragg's angles for intense peaks. Small concentrations of crystals and their small sizes do not allow to accurately determining the size. An approximate estimation according to the Scherer formula gives a size of approximately 5-6 nm.

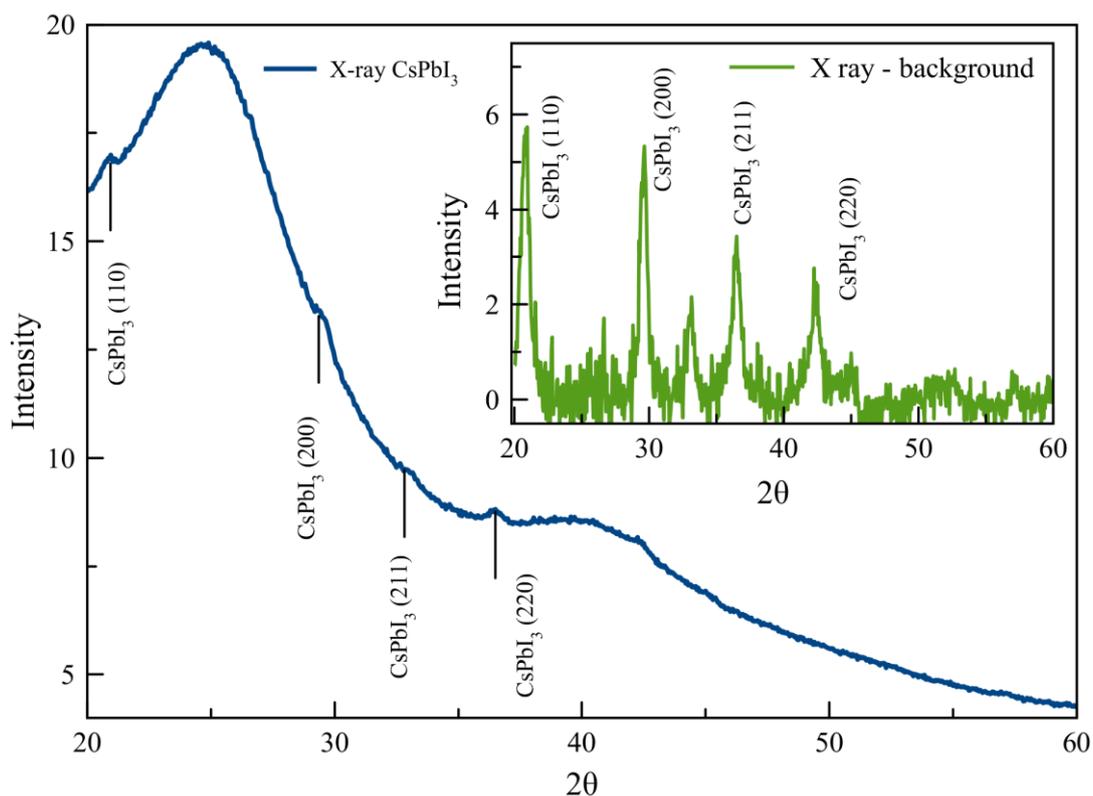

Figure3  X-ray diffraction patterns of the glasses doped with CsPbl3 (JCPDS No 80-4039); Inset shows X-ray diffraction patterns after subtraction the background.

Figure 3 shows of the X-ray diffraction patterns of the glasses doped with $CsPbI_3$ prepared by the second method. According to [12], a certain difficulties occurs obtaining a cubic modification of $CsPbI_3$. It is well known that bulk $CsPbI_3$ crystallize in orthorhombic, tetragonal, and cubic polymorphs of the perovskite lattice with the cubic phase being the high-temperature state as for all compounds $CsPbX_3$. At room temperature, there are non-luminescent, wider-bandgap orthorhombic phases $CsPbI_3$. $CsPbI_3$ is a highly luminescent and red in its three-dimensional cubic phase, yet yellow and non-luminescent upon conversion into its orthorhombic polymorph. When obtaining perovskite QDs using the traditional first method, a yellow phase is released and NCs sizes increase when the heat treatment duration increases. Only after reaching the orthorhombic crystals of larger sizes, they transform into the cubic phase. Thus the conversion of the already formed large orthorhombic NCs into the cubic black NCs takes a place as in [29] and cubic NCs are preserved during subsequent cooling of the glass. Therefore, the first method is not possible to form a cubic modification of small size QDs. The data presented in Figure 3 relate to the production of QDs by the second method, when the formation of QDs occurs during the cooling of the supersaturated melt. Small concentrations of crystals and their small sizes do not allow to accurately determining the size. An approximate estimate according to the Scherer formula gives a size of approximately 5-6 nm, what is significantly smaller than the radius of the Bohr exciton

## 3.2 Luminescent properties

It was shown, that for colloidal perovskite nanocrystals of $CsPbX_3$ with orthorhombic (X= Br) and cubic crystal structure (X=Cl, I), the bright PL (with quantum yields of 10 −80%, the lowest values for $CsPbCl_3$) is retained in anion-exchanged $CsPbX_3$ NCs. The peak widths (full width at half maximum - FWHM)  were ranging from 12 nm for $CsPbCl_3$ to 40 nm for $CsPbI_3$ [ 3]. Note that the exciton Bohr radius determining the range of size confinement for the crystals under study are chloride (5 nm), bromide (7 nm), and iodide (12 nm), respectively. "Black" phase of $CsPbI_3$ is metastable in the family of $CsPbX_3$

### 3.2.1 Perovskite nanocrystals in system $CsPb(Cl_xBr_{(1-x)})_3$

To compare the luminescence parameters of colloidal perovskites and perovskites in FP glass, we considered a $CsPb(Cl_xBr_{(1-x)})_3$ system with a change in the ratio of the levels of bromides and chlorides. The luminescence spectrum of glass upon the introduction of an activator of $NaCl/BaBr_2=5$ is shown in Figure 4a (X-ray diffraction patterns of this glass sample are shown in Figure 1). The maximum of the emission band is corresponds to wavelength 448 nm. This means that the size of the $CsPbCl_3$ nanocrystals exceeding the exciton Bohr radius.

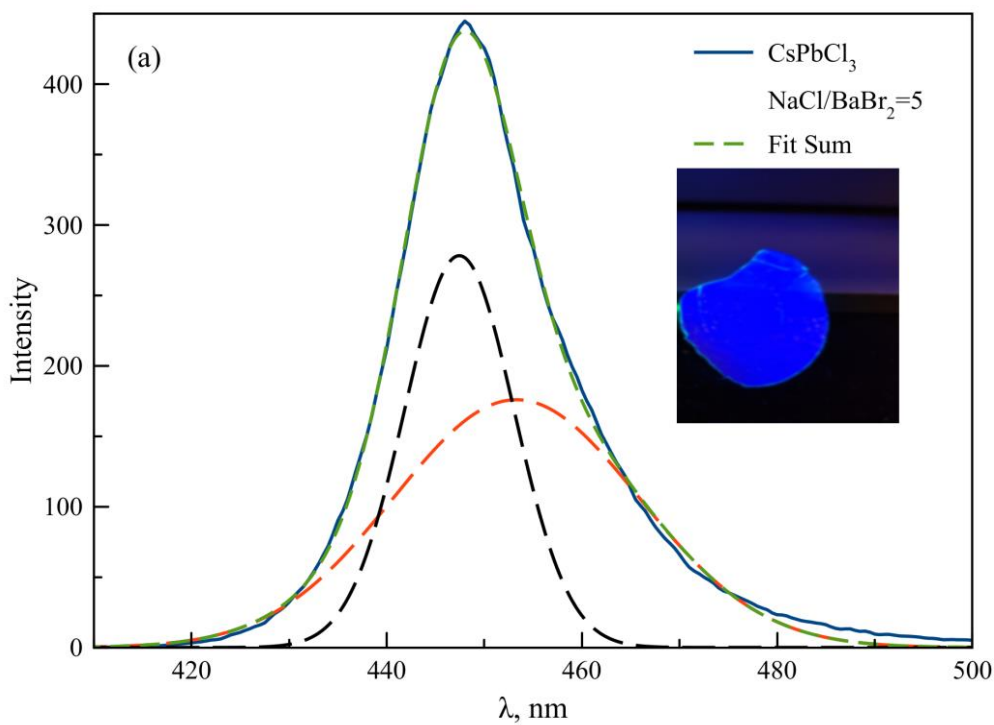

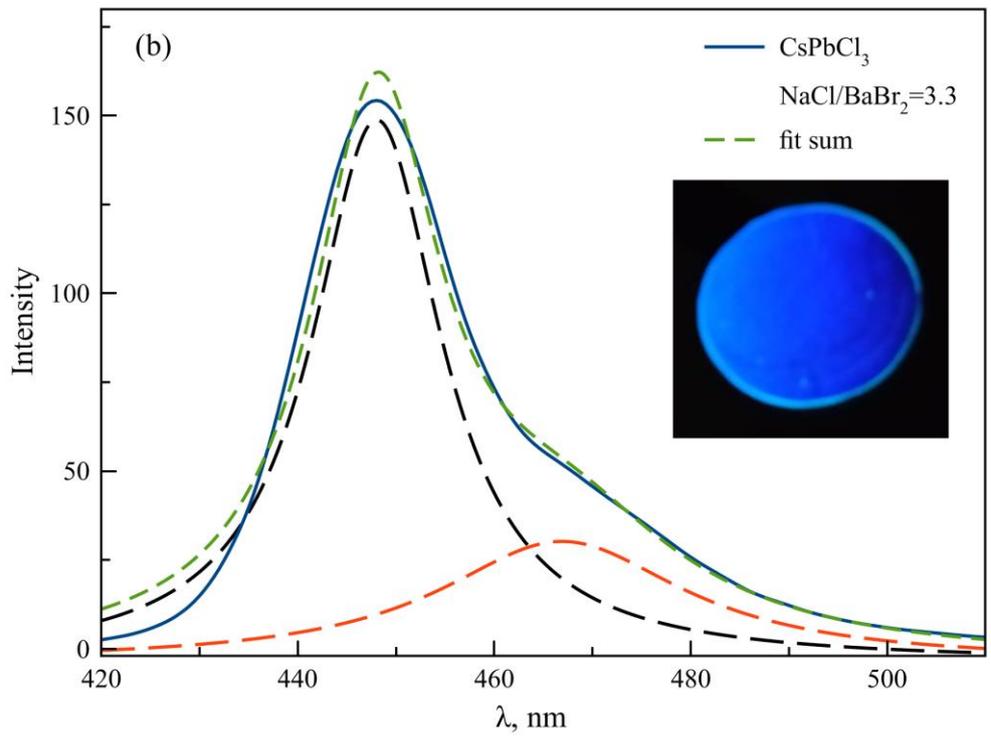

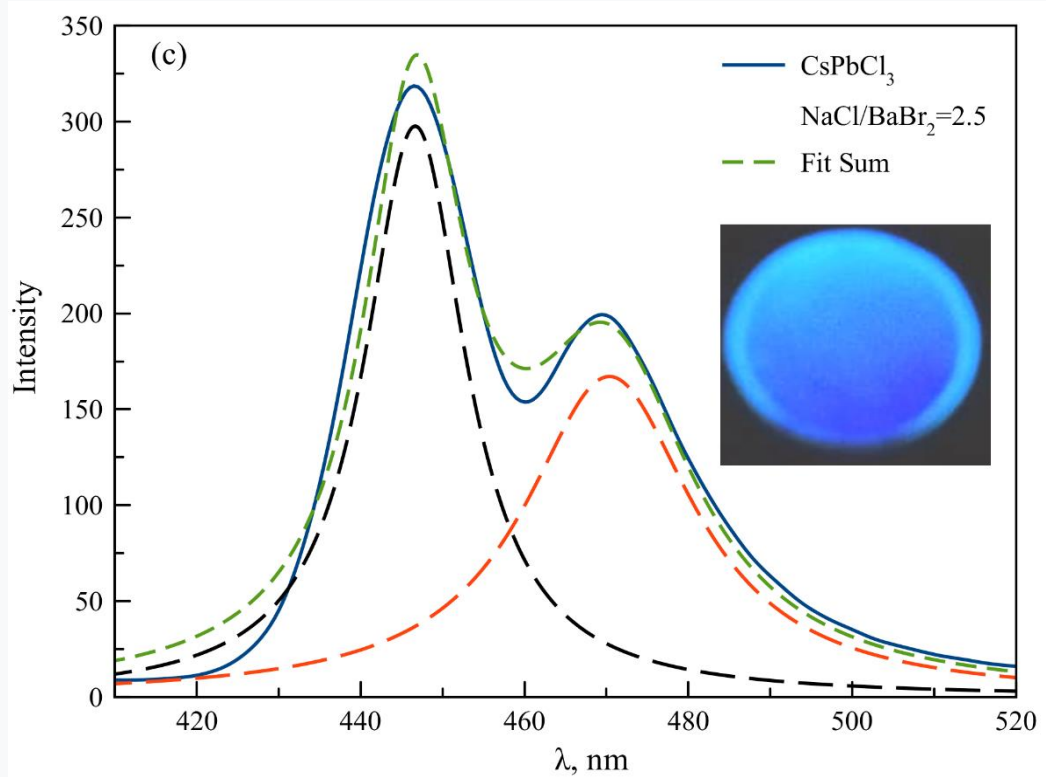

Figure 4 Glass luminescence spectra upon the introduction of an activator in the ratio NaCl/BaBr$_2$=5 (a); NaCl/BaBr$_2$=3.3 (b); NaCl/BaBr$_2$=2.5 (c).

Decrease of the ratio of the sodium chloride to bromine content to 3.3 results in an appearance of additional shoulder at the long-wavelength edge of the luminescence band (see Figure 4b). The decomposition of the PL band into two Lorentzian contours allowed us to quite accurately determine the positions of the maxima. With a NaCl/BaBr$_2$ =3.3, two bands appear: $\lambda_1$=448 nm and $\lambda_2$=460 nm with FWHM of the first PL peak 16 nm and the second 40 nm. Analyzing the obtained values using the results of [37], it can be assumed that the first band belongs to large CsPbCl$_3$QDs (d=7 nm), and the second one shows the appearance of small (<3.5 nm) CsPbBr$_3$QDs. A further increase in the bromine content in the glass led to a sequential increase in the intensity of the perovskite bromide band (see Figure 4c). The second PLmaximum was shifted to 470 nm, which corresponds to a QDs diameter of 3.7 nm [37], while the band corresponding to chloride NCs maintained its position at 448 nm. FWHM of the band of CsPbCl$_3$ QDs was 14nm, the FWHM of CsPbBr$_3$QDs was 24 nm, which corresponds to the characteristic values of these colloidal NCs.

Such a sequence of the luminescence spectra changes in depending on the ratio of chlorine and bromine testified in favor of our hypothesis about the formation of a crystallization nucleus in the form of CsPbBr$_3$ QDs and the catalyzed crystallization of CsPbCl$_3$. With an increase in the bromide content in the glass, in addition to heterogeneous crystallization of theCsPbCl$_3$ NCs, CsPbBr3 QDs simultaneously formed. Confirmation of this hypothesis was the coincidence of the FWHM values of the luminescence band of chloride NCs (12-14 nm) and bromide QDs (20 nm) with the values of colloidal NCs at the room temperature.

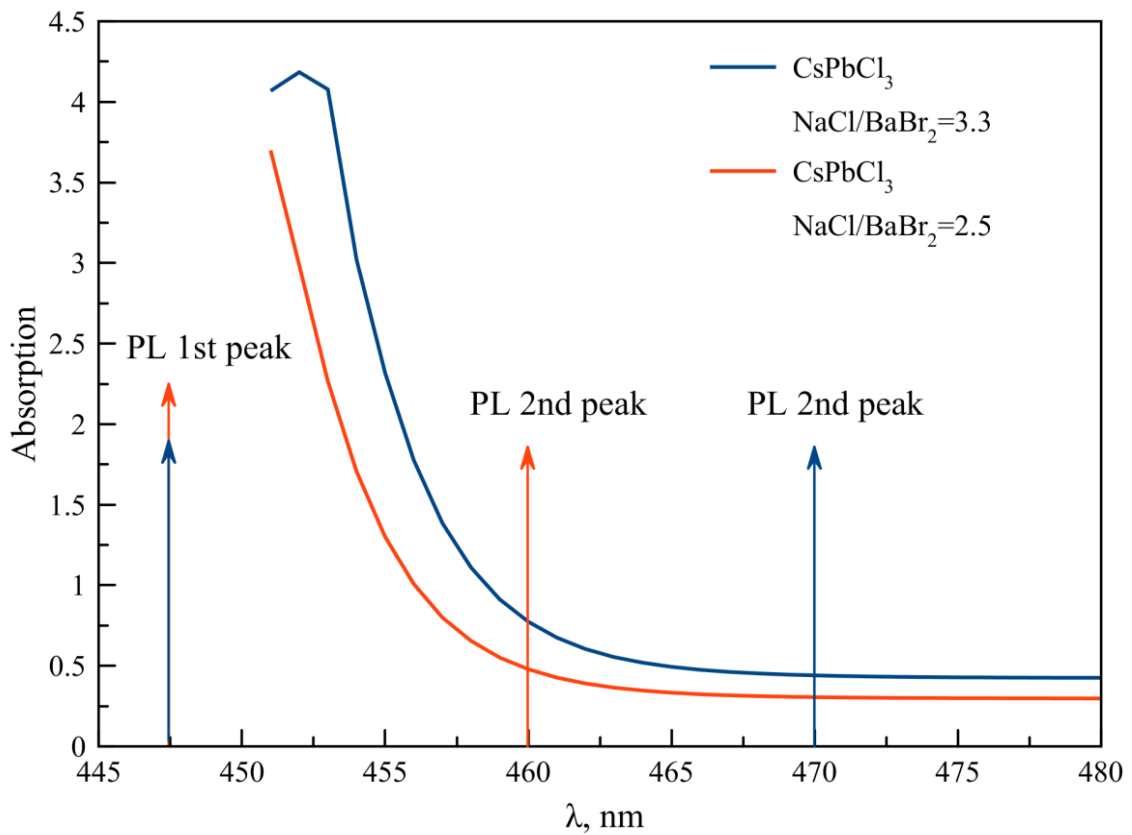

Figure 5. Absorption spectra of glasses upon the introduction of an activator in the ratio NaCl/BaBr$_2$ = 3.3 (blue line) and NaCl/BaBr$_2$ = 2.5 (red line). The arrows indicate the positions of the luminescence peaks of glasses.

An additional argument in favor of the version of heterogeneous crystallization of chlorine nanocrystals is the absorption spectra of glasses (Figure 5). The Figure 5 shows the position of the luminescence maximum which designated as PL1 (CsPbCl$_3$) and PL2 (CsPbBr$_3$) relative to the band gap.

An analysis of the absorption and luminescent spectra leads to the following conclusions: (i) increase of the bromine content results in increase of the bromine QDs size (a absorption edge and luminescent band are shifting to the higher wavelengths); (ii) the PL band of CsPbCl$_3$ NCs does not change position and is shifted relatively to the band gap towards lower wavelength, because band gap position is determined by CsPbBr$_3$ QDs. It is difficult to obtain the correct values of the quantum yield for the chlorine NCs due to high absorption in the range of the PL band. For the series of glasses under study, the quantum yield was obtained 45-50%.

We assume that the reason for the difficulties to obtain mixed perovskite nanocrystals in FP glasses lies in the formation of chloride and bromide nanocrystals in different polymorph forms. Orthorhombic CsPbBr$_3$ NCs and the cubic CsPbCl$_3$ NCs are formed in the FP glass.

### 3.1.2 CsPbBr$_3$ QDs

In the large variety of perovskite materials, CsPbBr$_3$ NCs have attracted most of attention since they demonstrate highest quantum efficiency and superior stability against photoexcitation and oxidation leading to outstanding optical properties. Due to the fact that CsPbBr3 has an orthorhombic modification in the nanocrystalline state, as in bulk, a simple method for producing NCs by grinding was proposed [14].

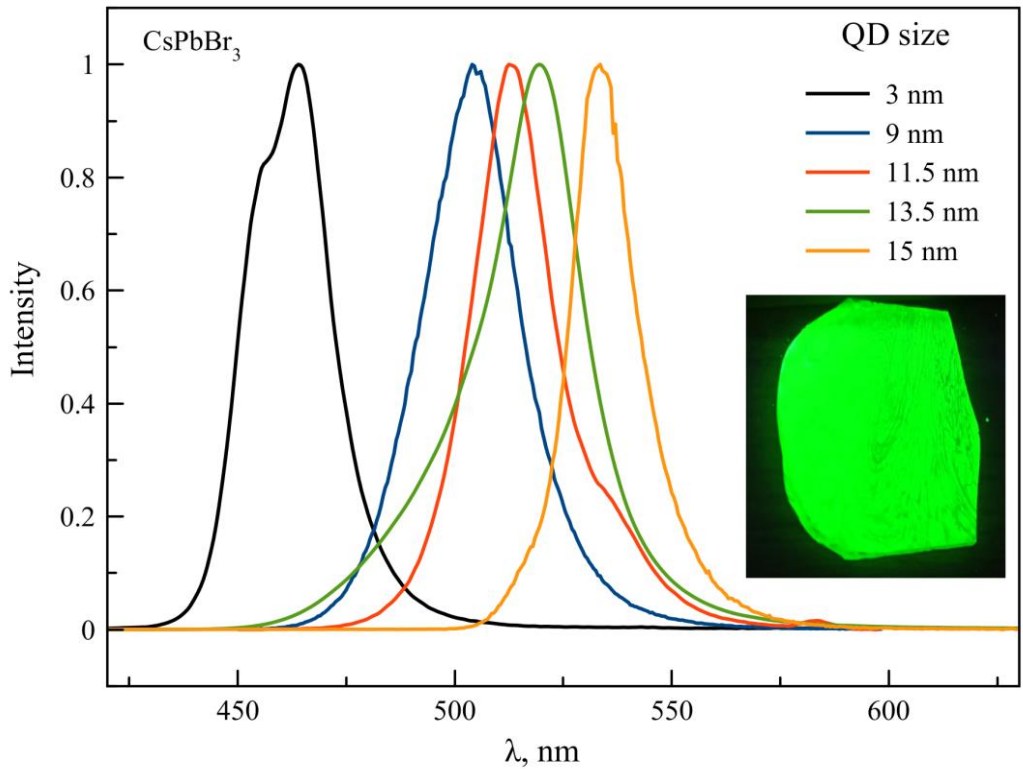

Figure 6    Luminescence spectra of the glasses doped with CsPbBr$_3$ QDs with sizes from 3.5 to 11 nm. QDs sizes are estimated based on a comparison with the data presented in [37]

Figure 6 demonstrates the PL spectrum of CsPbBr$_3$ QDs for different QDs sizes. The PL band shifted to the higher wavelength with increasing of the sizes of QDs in FP glasses containing only bromine. Depending on the QD size, the PL band shifted from 460 nm to 536 nm. According to [37], with increasing QDs sizes from 3.5 to 11.5 nm, a band shift should be observed from 470 to 515 nm. Considering that the band gap of CsPbBr$_3$ crystals is 525 nm at room temperature, the band 536 nm corresponds to the luminescence of a nanocrystal with sizes more than two times Bohr exciton radius. It should also take into account the effect of the matrix with an increase in the size of halide nanocrystals [28]. Intriguingly, we observed that two luminescence bands corresponding to two different sizes (two-mode crystallization) can occur simultaneously, for example, the curves PL for the size 3.5 nm (black line) and 11.5 nm (red line) in Figure 6. An explanation of this phenomenon is not assumed in this paper. For this series of glasses, the quantum yield obtained 50-55%. CsPbBr$_3$ NCs can be promising as spectrally narrow green primary emitters in backlighting of liquid-crystal displays.

### 3.1.3  CsPbI$_3$ QDs

Cesium lead iodide (CsPbI$_3$) in cubic phase (α phase) is the mostly desired light harvester in solar cells [38]. The bandgap of the CsPbI$_3$ NCs is around 1.73 eV and PL spectrum is shifted up to 700 nm. However, cubic CsPbI3 can only keep stable at high temperature of above 300 °C [38]. As temperature decreases, CsPbI3 suffers from thermodynamically phase transition to undesired orthorhombic phase (δ phase) with a wide bandgap of 2.82 eV.

Colloidal CsPbI3 NCs have a poor stability due to this an undesired phase-transition can occur. Photoactive high-temperate black α-phase rapidly degrade into photoinactive yellow δ-phase when exposed to air. The features of the NCs formation in FP glasses were formulated in Section 3.13.

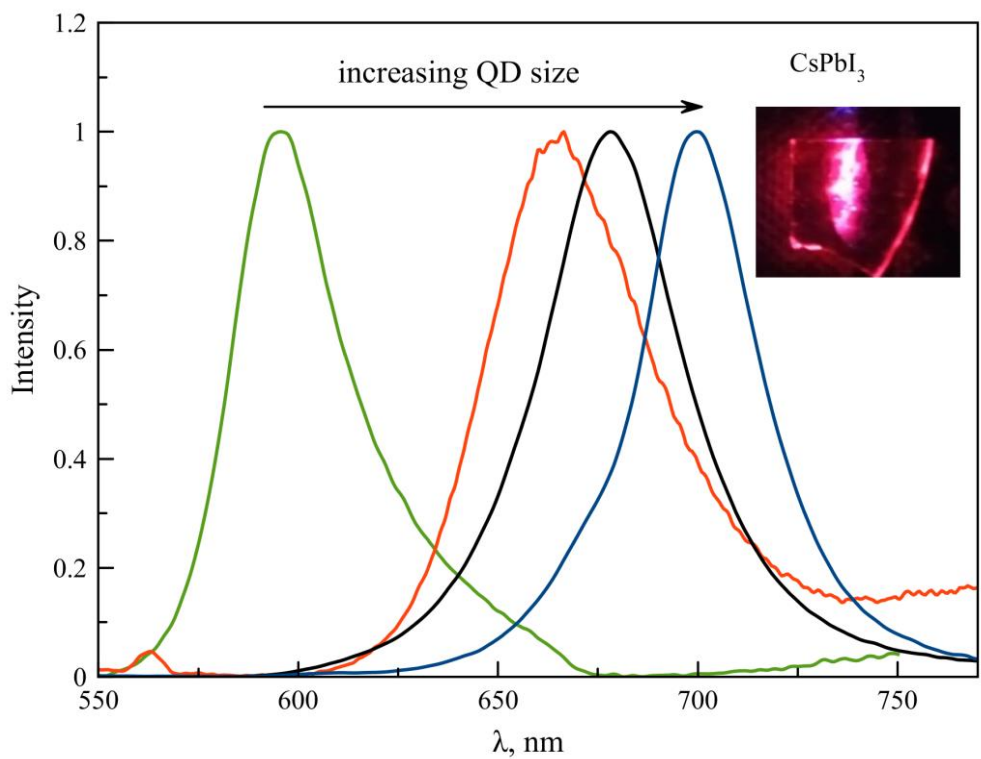

Figure 7    Luminescence of CsPbI3 QDs with increasing size;

The Figure 7 shows the PL for different sizes of CsPbI3 QDs. Changing the QDs average size led to a shift of the luminescence band by 100 nm. Large QDs with PL maxima of 670–700 nm are stably formed by both synthesis methods. The smallest sizes are formed only by the second method during cooling of the melt. In latter case, apparently, only the cubic phase appears, as for copper halides [29].

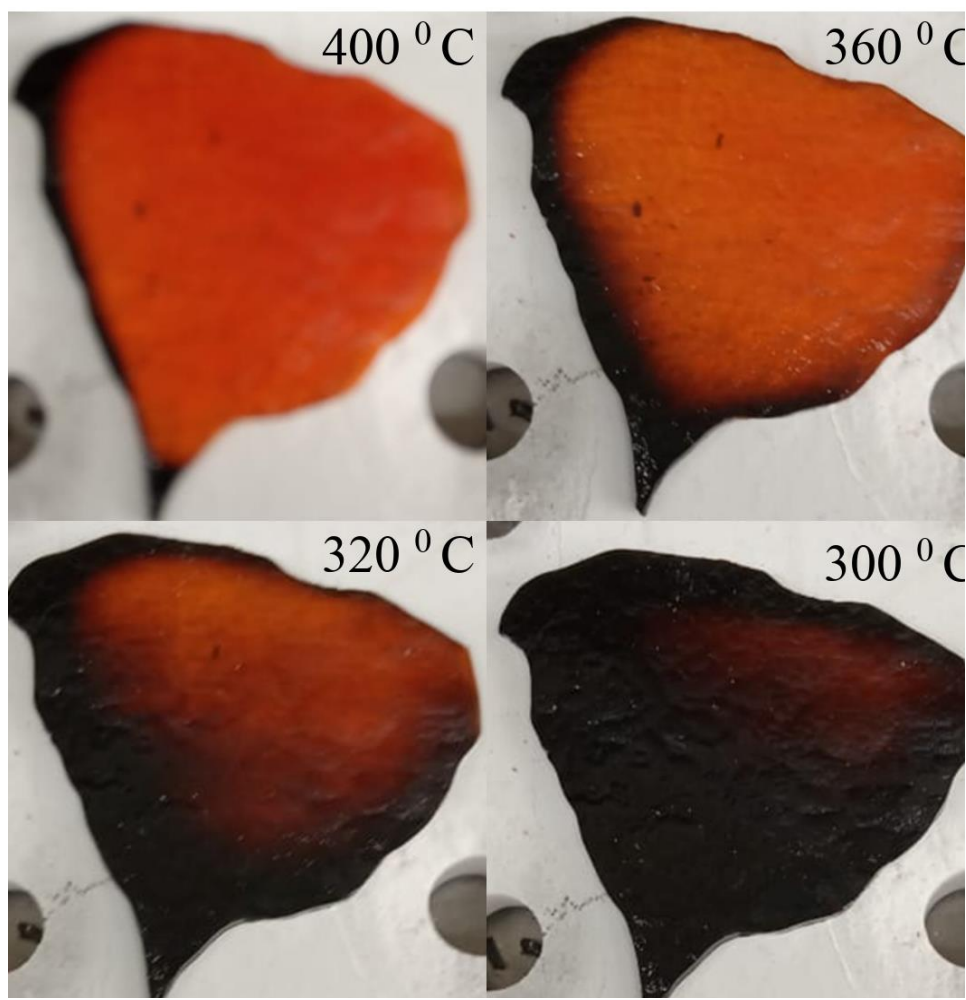

Figure 8 Coloration of glass with CsPbI$_3$ during spontaneous cooling from 400ºC to 300 ºC

Figure 8 shows the color change of CsPbI$_3$ glass upon cooling from 400 to 300 °C. The observed effect can be associated both with a temperature shift of the band edge and with a phase modification of the NCs.

To stabilize the crystal structure of α-phase CsPbI$_3$ and to lower the bandgap of CsPbBr$_3$, researchers started to focus on I/Br mixed halide CsPb(I$_x$Br$_{(1-x)}$)$_3$. The stable colloidal CsPb(I$_x$Br$_{(1-x)}$)$_3$ (0.67≤x≤1) NCs were developed [38, 39]. The preparation of mixed CsPb(I$_x$Br$_{(1-x)}$)$_3$ nanocrystals in glasses has been shown [12].

We also made an attempt to obtain mixed CsPb(I$_x$Br$_{(1-x)}$)$_3$ NCs in FP glass. The attempt was unsuccessful (Figure 9). Two PL bands were observed in the luminescence spectrum of the glass doped with BaBr$_2$/BaI$_2$=2. We suppose the first band was due to the growth of CsPbBr$_3$ QDs. The second band can be associated with both small CsPbI$_3$ QDs and mixed QDs. The green luminescence band is shifted to the region of higher energies in comparison with the band gap. This result confirms the difficulties in the formation of perovskites with a mixed anion in fluorophosphate glasses

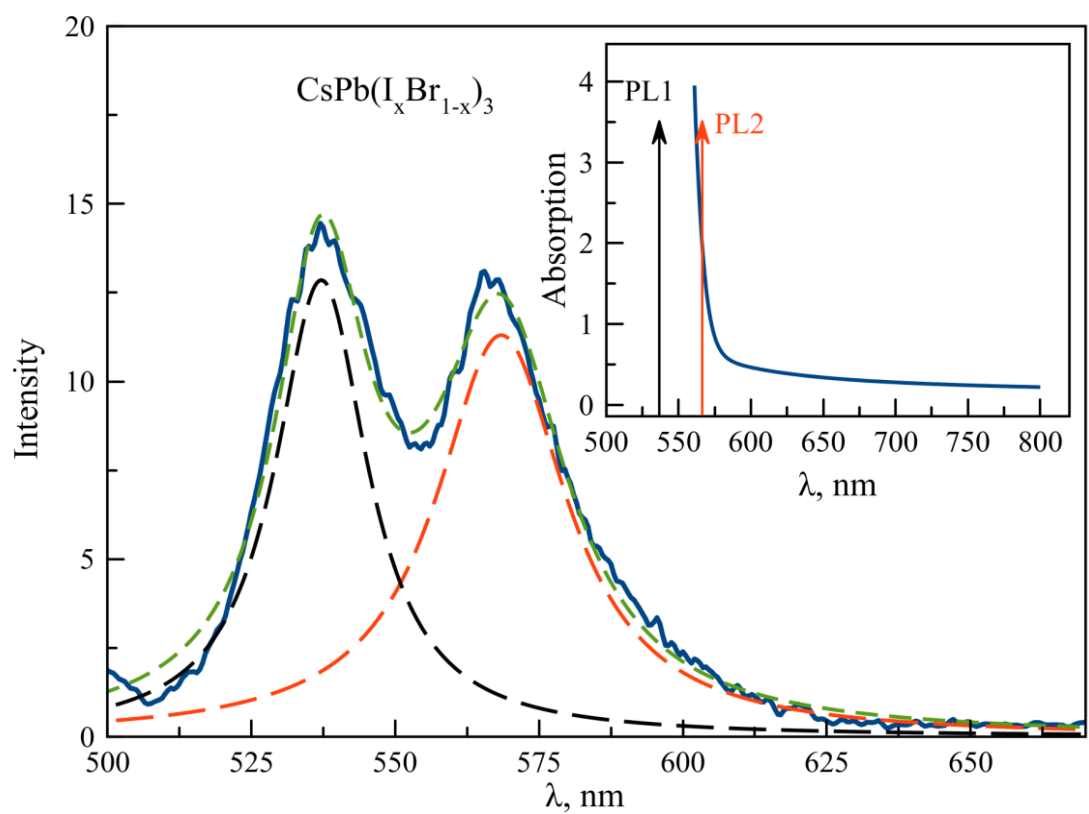

Figure 9 Glass luminescence spectra upon the introduction of an activator in the ratio $BaBr_2/BaI_2=1$. Inset adsorption of the glass, PL1 — $CsPb(Br)_3$ luminescent band position, PL2— $CsPb(I)_3$ luminescent band position.

## 4. Summary

$CsPbX_3$ (Cl, Br, I) QDs with tunable composition were precipitated in FP glasses through conventional melt-quenching method. Photoluminescence of $CsPbX_3$ quantum dots were tuned from 447 to 700 nm due to effect of quantum confinement and exhibit high luminescent intensity with quantum yield ~50% in spectral region 450-550 nm and narrow photoluminescence band emission. The combined introduction of two anions (Cl/Br and Br/I) led to the appearance of two characteristic luminescence bands, which indicates the formation of two types of quantum dots, and proves the difficulty of the anion substitution process. It should be noted that repeated heating of $CsPbX_3$ QDs (Cl, Br, I) in glasses up to 400 °C does not lead to degradation of luminescent properties.

**Acknowledgements**

This work was funded by Russian Science Foundation (Grant № 19-13-00343). M.S.K. thank the St. Petersburg State University Research Grant No. 51125686.